\documentclass[11pt, draftclsnofoot, onecolumn]{IEEEtran}

\usepackage{cite,graphicx,amsmath,amssymb}
\usepackage{subfigure}
\usepackage{fancyhdr}
\usepackage{mdwmath}
\usepackage{mdwtab}
\usepackage{balance}
\usepackage{xcolor}
\usepackage{bm}
\usepackage{amsthm}
\usepackage{algorithm}
\usepackage{algorithmic}
\usepackage{multirow}
\usepackage{flafter}

\begin{document}
\title{User Association and Resource Allocation in Unified NOMA Enabled Heterogeneous Ultra Dense Networks}

\author{
\IEEEauthorblockN{Zhijin~Qin,~\IEEEmembership{Member,~IEEE,}
Xinwei Yue,~\IEEEmembership{Student Member,~IEEE,}
 Yuanwei~Liu,~\IEEEmembership{Member,~IEEE,}
 Zhiguo~Ding,~\IEEEmembership{Senior Member,~IEEE,}
and Arumugam~Nallanathan,~\IEEEmembership{Fellow,~IEEE}
 }
 \thanks{Z. Qin and Z. Ding are with Lancaster University, Lancaster, UK, LA1 4YW, email: \{zhijin.qin, z.ding\}@lancaster.ac.uk.}
\thanks{X. Yue is with Beihang University, Beijing 100191, China, email: xinwei\_yue@buaa.edu.cn.}
\thanks{Y. Liu and A. Nallanathan are with Queen Mary University of London, London, UK, E1 4NS, email: \{yuanwei.liu, a.nallanathan\}@qmul.ac.uk.}

}

\maketitle

\begin{abstract}
Heterogeneous ultra dense networks (HUDNs) and non-orthogonal multiple access (NOMA) have been identified as two proposing techniques for the fifth generation (5G) mobile communication systems due to their great capabilities to enhance spectrum efficiency. This article investigates the application of NOMA techniques in HUDNs to support massive connectivity in 5G systems. Particularly, a unified NOMA framework is proposed, including power-domain NOMA and code-domain NOMA, which can be configured flexibly to serve different applications scenarios. As a further advance, the unified NOMA framework enabled HUDNs is further investigated, with particular focuses on the user association and resource allocation. Two case studies are provided for demonstrating the effectiveness of the unified NOMA enabled HUDNs. Finally, some main challenges and promising research directions in NOMA enabled HUDNs  are identified.
\end{abstract}


\section{Introduction}
The last decade has witnessed the densification of wireless networks due to the various types of wireless communication services. In order to support  explosive data traffic, the concept of heterogeneous network (HetNet) has been proposed by overlaying small cells with low transmit power to macro cells. Through the dense deployment of small cells,  throughput and spectrum efficiency of cellular networks can be enhanced significantly~\cite{Zhang:2016,Ge20155G,Zhang:2017}. Moreover, it is predicated that Internet of Things (IoT) will bring critical challenges for the fifth generation (5G) communication systems  as billions of devices  are to be connected. In order to support  massive connectivity with heterogeneous quality of service (QoS),  non-orthogonal multiple access (NOMA) has attracted extensive attentions  due to its potential capability to enhance spectrum efficiency~\cite{Saito:2013,Dai:2015,Ding2015Application,Islam:2017}. The key idea of NOMA is to enable multi-user transmission within the same resource block (RB), i.e., frequency/time, by using various power levels and/or different codes. Driven by the key characters of heterogeneous ultra dense networks (HUDNs) and NOMA, it is natural to invoke NOMA technique in HUNDs to support heavy data traffic as well as  provide massive connectivity.

Existing NOMA can be mainly categorized into power-domain NOMA (PD-NOMA) and code-domain NOMA (CD-NOMA) including low-density spreading CDMA (LDS-CDMA), low-density
spreading OFDM (LDS-OFDM), and sparse code multiple access (SCMA), which distinguish users by different power levels and codes, respectively. Despite of the growing attempts and extensive efforts on NOMA, most of the studies have focused on the performance analysis of various NOMA techniques individually, such as PD-NOMA and CD-NOMA. However, different scenarios have different preferred NOMA techniques. For example, if users experience very bad channel conditions due to the near-far effect or in a moving network, PD-NOMA can be a better candidate. If users experience poor channel conditions but requiring high reliability, SCMA is preferred due to its shaping gain and near-optimal message passing algorithm (MPA) detection. Therefore, it is desired to design a unified NOMA framework for 5G systems to support various scenarios. The core idea of the proposed unified NOMA is to provide a multiple access (MA) framework,  which is capable of supporting massive connectivity with heterogeneous QoS by using the same hardware infrastructure.

The goal of this article is to provide an unified NOMA framework and investigate its application in HUNDs to support massive connectivity for 5G and IoT. As both the dense deployment of small cells and the non-orthogonality in resource sharing bring severe interference, user association and resource allocation are very challengeable to support massive connectivity. Particularly, the following two issues should be addressed when designing the unified NOMA enabled HUNDs:
\begin{enumerate}
  \item \textbf{User association}: user association process should consider both  intra interference from the same cell  and inter interference introduced by the same cell as well as neighboring cells. Therefore, controlling the number of users assigned to each cell can be an efficient approach to control interference.
\item \textbf{Resource allocation}: once users are allocated into different cells, how to assign them with the most suitable cell and proper transmit power for NOMA users within the same cell becomes critical. Thus, efficient resource allocation and interference control schemes are more than desired in NOMA enabled HUNDs.
\end{enumerate}

The rest of this article is organized as follows. We first introduce the concepts of HUNDs and NOMA techniques. Then we explore how NOMA techniques can be applied in HUNDs to enhance spectrum efficiency and support massive connectivity. Subsequently, we propose a unified NOMA framework and investigate its application in HUNDs in section III. Specifically, we provide both the uplink and downlink cases for the unified NOMA enabled HUNDs. We discuss user association and resource allocation in NOMA enabled HUNDs. In section IV, we provide  related case studies for the proof of concept. We also identify some potential research challenges in unified NOMA enabled HUNDs in Section V before giving the conclusion remarks in Section VI.

\section{Overview of NOMA-Enabled HUDNs}
In this section, we first introduce the basic principles of HUDNs and NOMA techniques, respectively. Then we discuss  the general architecture of  NOMA enabled HUDNs to support massive connectivity.

\subsection{HUDNs}
The density of wireless networks is invoked by the large number of devices, such as tablets, smart phones, and IoT devices. To provide higher throughput and spectrum efficiency, HUDN technique has attracted extensive research interest~\cite{Zhang:2016}. Particularly, HUDNs refer to networks that involve many different types of small cells to make the access points getting as close as possible to end users \cite{Ge20155G}. Besides macro cells,  HUDNs contain cells with various sizes, such as pico cells, femto cells, and relays, which normally transmit at lower power than macro cells and can offload data traffic from the macro cells.  With the increasing density of small cells, the backhaul network capacity and spectrum efficiency can be enhanced significantly. However, it is unrealistic to deploy small cells with infinite density in practical scenarios. Therefore, extensive research efforts are required to capture the reality of densification networks.

\subsection{NOMA}

In order to satisfy the requirements of massive connectivity and higher spectrum efficiency, NOMA has been identified as a proposing technique in 5G systems~\cite{Saito:2013,Dai:2015,Ding2015Application,Islam:2017}. Compared with the orthogonal multiple access (OMA) techniques, such as FDMA and TDMA, NOMA breaks the orthogonality by allowing multiple users sharing the same physical resource. More particularly,  the virtue of superposition coding is adopted to generate signals for multiple users at transmitters. At receivers, according to the adopted MA approaches, different multi-user detection (MUD) algorithms, such as successive interference cancellation (SIC) and MPAs, can be implemented to remove co-channel interference. In general, MA techniques can be classified into PD-NOMA and CD-NOMA. Extensive research work has been carried out on PD-NOMA~\cite{Saito:2013,Ding2015Application,Sun2017TCOM,Liu2017HetNets,Fang:2016,jingjing2016Hetnets} and CD-NOMA~\cite{Nikopour2013Sparse,ChenPattern7526461,Yuan2016Multi}, respectively.

\subsection{Understanding NOMA in HUDNs}
Driven by the above overview,  both HUDNs and NOMA are considered as promising techniques in 5G systems to  enhance spectrum efficiency and to support massive connectivity. Therefore, NOMA enabled HUDNs are capable of further improving the spectral efficiency by offering more access opportunities.

\begin{figure}[!t]
    \begin{center}
        \includegraphics[width=6.5 in,  height= 3.1 in]{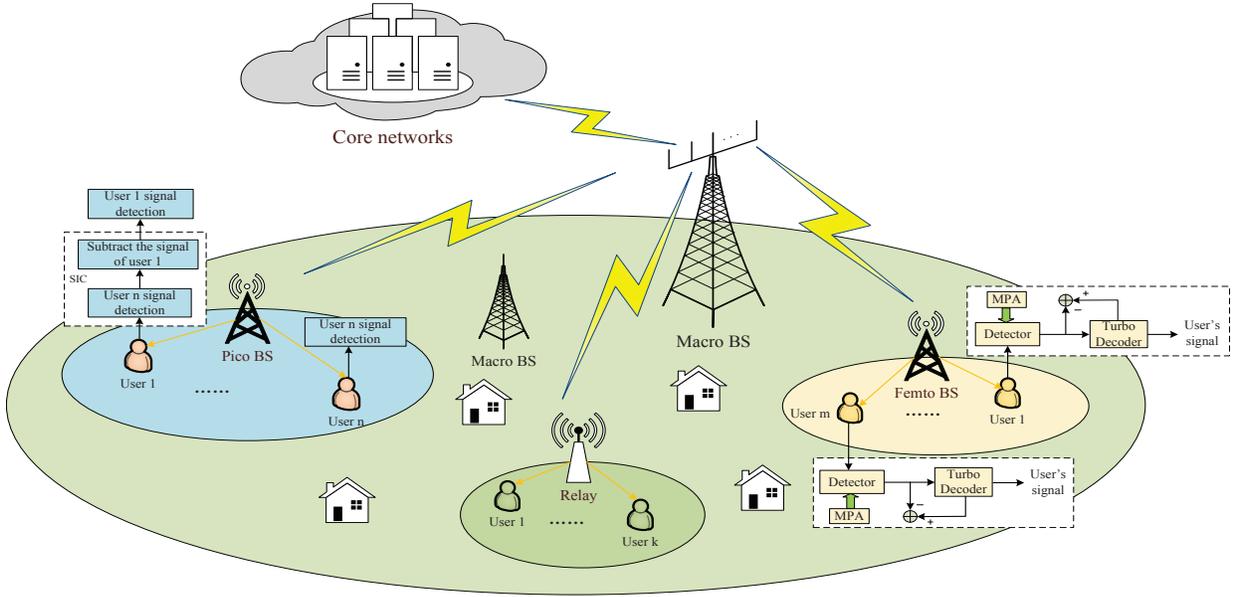}
        \caption{Illustration of the NOMA enabled HUDNs.}
        \label{Mg_system_model}
    \end{center}
\end{figure}

Fig.~\ref{Mg_system_model} illustrates the architecture of NOMA enabled HUDNs. It can be observed that  small cells, including femto cells, pico cells, and relays, are densely deployed through the whole network, which can enhance the system capacity significantly. In NOMA enabled HUDNs, NOMA technique is adopted in each small cell to enable multiple users sharing the same RB, while the massive multiple-input multiple-output (MIMO) technique is employed by macro cells. More particularly,  macro cells can be connected to core networks by optical fiber or wireless backhaul networks, and it is assumed that the number of antennas equipped at each macro BS is much larger than the number of users.  For small cells, each user is equipped with single antenna and NOMA techniques are adopted to support multi-user transmission over the same RB.

More specifically, the pico cell in Fig.~\ref{Mg_system_model} gives an example of PD-NOMA, which has low-complexity receivers  and is preferred by applications with less restriction on reliability. It can be observed that pico BS allocates different power levels for PD-NOMA users according to their channel conditions. At User 1, who has best channel condition and lowest transmit power, by applying SIC, signals for the other users with higher transmit power levels will be detected first  and abstracted from the received signal. Then the desired signal for User 1 can be obtained. While for User $n$ assigned with highest transmit power, signals  for other users will be treated as noise when detecting its own signal. Moreover, femto cell shown in Fig.~\ref{Mg_system_model} gives an illustration of multi-user transmission by invoking CD-NOMA, which is more suitable for cases requiring higher reliability. We observe that  users in femto cell carry out MUD-based MPA individually to alleviate error propagation effects. After MPA detection, the soft information of users is output to Turbo decoder, and the iterative process between MPA detector and Turbo decoder  further enhances the detection performance.

In HUDNs,  users may experience very different channel conditions when connecting to BSs from different tiers. It has been identified that different NOMA techniques have different performance in terms of supporting massive connectivity under various scenarios, i.e., applications requiring different   reliabilities and/or transmission rates. The same user may need different NOMA techniques when it experiences various channel conditions and has different transmission requirements. Therefore, different hardware architectures are required to support such various scenarios, which brings bottleneck for the real implementation of NOMA techniques. It is necessary to provide a unified NOMA framework for HUNDs, which can be implemented on the same hardware architecture but with flexible capability to support various scenarios.

\section{A Unified NOMA Framework for HUDNs}
In this section, we first propose a unified NOMA framework for HUDNs, in which both the uplink and the downlink are investigated, respectively. Then we address user association and resource allocation issues in the considered HUDNs with the proposed unified NOMA framework.

\subsection{Overview of HUDNs with Unified NOMA}
\begin{figure}[!t]
\centering
\subfigure[Uplink NOMA]{
\label{Fig.sub.1}
\includegraphics[width=3.0 in]{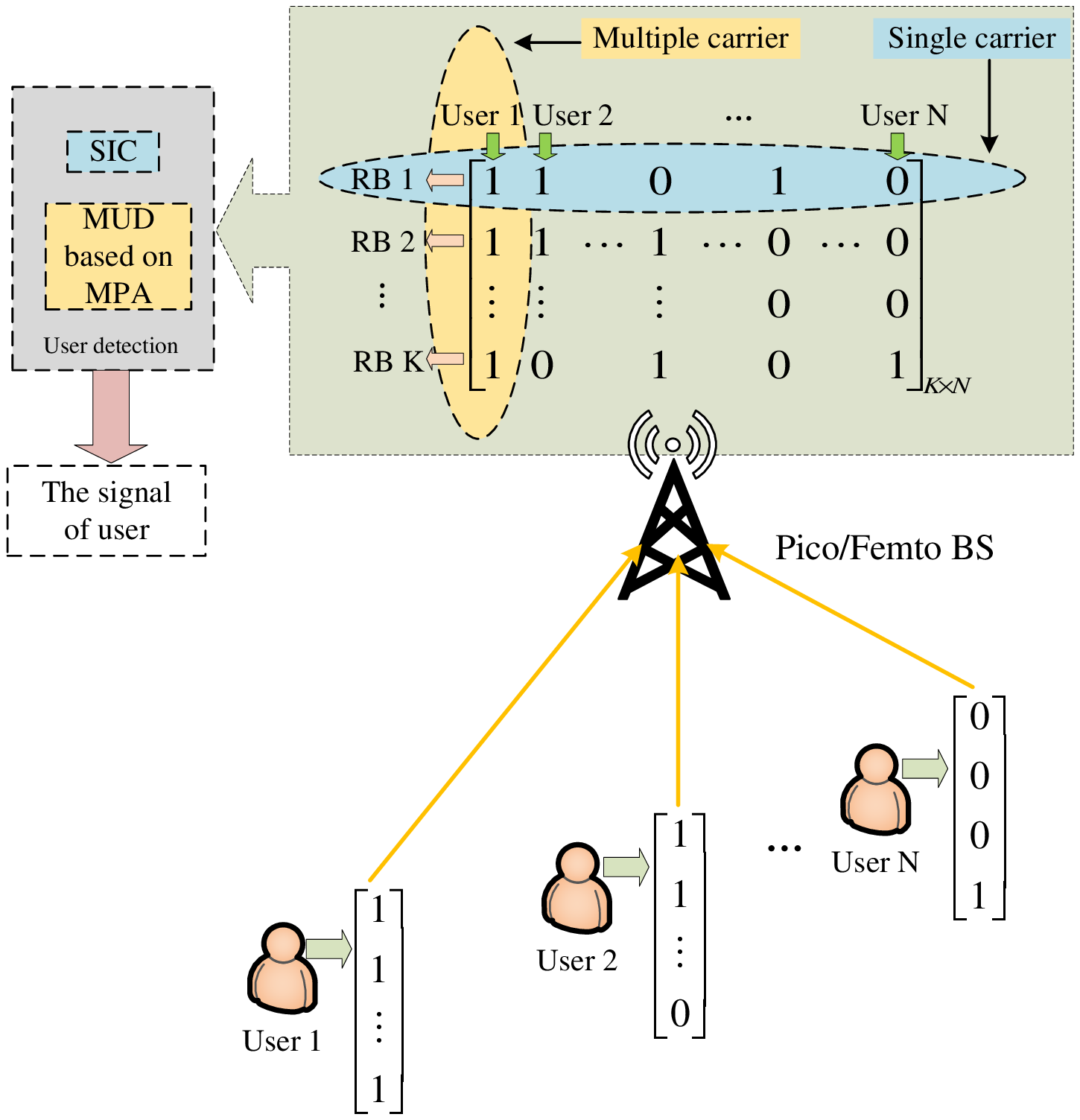}}
\;\;\;\;\;\;\subfigure[Downlink NOMA]{
\label{Fig.sub.2}
\includegraphics[width=3.0 in]{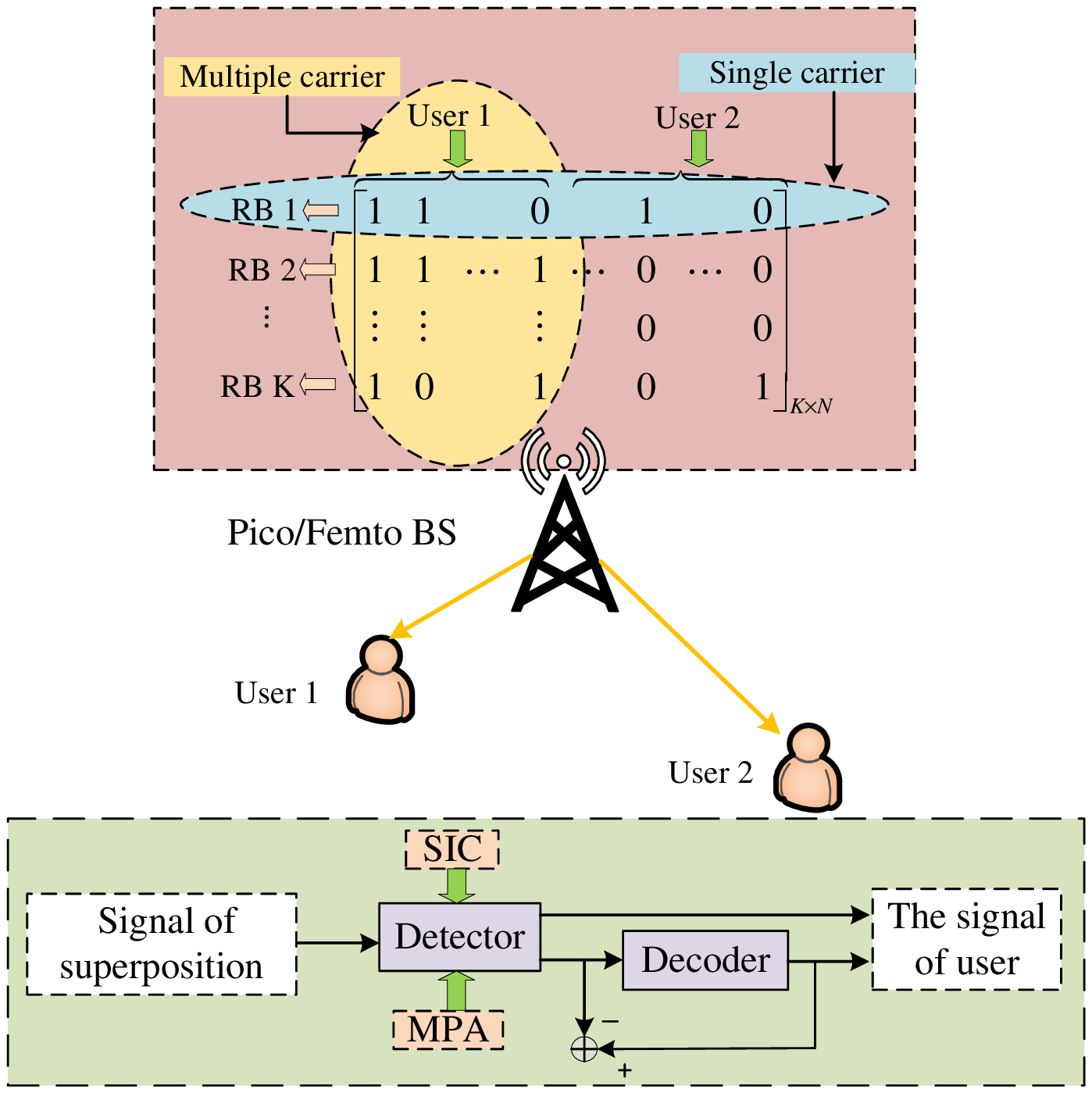}}
\caption{Uplink and downlink NOMA systems}
\label{Fig. 1}
\end{figure}

In this part, we propose a unified NOMA framework, which contains both PD-NOMA and CD-NOMA techniques. As shown in Fig.~\ref{Fig. 1}, we first map the superposed signals of multiple users to single RB or multi-RB over a sparse matrix, in which most of the elements are zero. Note that single RB and multi-RB correspond to single carrier and multi-carrier, respectively.
In other words, single carrier NOMA (PD-NOMA) is the special case of multi-carrier NOMA (CD-NOMA).
The rows and columns of the sparse matrix represent different RB and different users, respectively. Here, ``1" represents the user occupies the corresponding RB and ``0" otherwise. The use of such a sparse matrix is essential to capture the features of SCMA~\cite{Nikopour2013Sparse} and pattern division multiple access (PDMA)~\cite{ChenPattern7526461}, where the optimal design of sparse matrix for CD-NOMA is capable of reducing detection complexity at receivers. In particular, SCMA and PDMA are belong to CD-NOMA's different types of forms, in which the equal and unequal column weight sparse matrixes are employed, respectively. It is worth noting that in HUDNs, each cell can choose PD-NOMA or CD-NOMA scheme based on the  pre-configuration of the BS. We will present the details of PD-NOMA and CD-NOMA, i.e., SCMA and PDMA, and use a two user case to illustrate how the unified NOMA framework works in the following.

Regarding PD-NOMA, a transmitter is capable of multiplexing multiple users via different power levels within the single subcarrier, i.e., the first row of sparse matrices illustrated in Fig.~\ref{Fig. 1}.  At receivers, PD-NOMA exploits SIC to remove the multi-user interference.  Additionally, PD-NOMA can be also realized in multi-carriers, with the aid of appropriate user scheduling and power allocation approaches \cite{Sun2017TCOM}. It is worth noting that downlink multi-user superposition transmission (MUST), which is essentially a special case of PD-NOMA, has been standardized.

CD-NOMA can be regarded as a special extension that directly maps data streams of multiple users into multiple carriers by using the sparse matrix or low density spreading code at transmitters, as shown in the multi-carrier case in Fig.~\ref{Fig. 1}. At receivers, multiple users are distinguished by MPA to obtain the multiplexing or coding gains. For example, SCMA utilizes a sparse matrix, in which each column can be selected from the predefined codebooks. With the multidimensional constellations to optimize codebooks, SCMA is able to achieve enhanced shaping and coding gains. While for PDMA, the core concept is to jointly optimize  transmitters with sparse pattern design and receivers with MPA-based detection. The design of sparse pattern can provide disparate diversity for multiple users and further reduce the complexity of detection. Additionally, phase shifting is an effective way to obtain constellation shaping gain. It is worth noting that the pivotal difference between these two schemes is that the number of RBs occupied by each user has to be the same in SCMA, while PDMA allows a variable number of RBs to be occupied by the same user. For another CD-NOMA scheme, muti-user shared access (MUSA) \cite{Yuan2016Multi}, each user's data symbols are spread by a special spread sequence to facilitate SIC implementation. This kind of spread sequence can be selected from the sparse matrix as illustrated in Fig.~\ref{Fig. 1}, which requires special design to achieve low cross-correlation. Note that multiple spreading sequences constitute a pool, where each user can select one sequence from it.

\subsection{Uplink and Downlink Design for NOMA Enabled HUDNs}

\subsubsection{Uplink Design}
To facilitate understanding the uplink design in the unified  NOMA framework, we present specific examples in the following. For uplink PD-NOMA, multiple users transmit messages to the BS by the same RB.  As shown in Fig. \ref{Fig.sub.1}, when using the sparse matrix for PD-NOMA, multiple users' signals are mapped into RBs in the first row of sparse matrix, while the rest rows of the sparse matrix are set to zeros. Notice that power control strategy is a vital issue in the uplink NOMA transmission, especially when powers received at users are significantly distinct. We employ SIC at BSs to decode and subtract the information of the nearby user first, then decode the message of the distant user. By doing so, the data rate of the distant user can be guaranteed. For CD-NOMA, each user selects one or several columns from the sparse matrix and then maps its information to multiple RBs by using the spreading code. For example, considering the case that each user selects only one column and spreads its message to ``1'', then the data streams of $N$ users are superposed through $K$ RBs to construct the sparse matrix with dimensions of $K \times N$ at the BS. The sparse properties of matrix is conducive for implementing the MPA detection. To guarantee the fairness among users, the nearby user selects one column with smaller column weight, while the distant user selects column with larger column weight. Such operation can enhance  reliability of received signals for multiple users.

\subsubsection{Downlink Design}
The key feature of the downlink design in the unified NOMA framework is that multiple data streams of different users are superimposed at BSs. Then BSs transmit the superimposed signals to multiple users simultaneously. More particularly, a BS maps multiple users' signals into single or multiple carriers over the sparse matrix. In the following, we use a two-user case to illustrate how the downlink of the unified  NOMA framework works.

For PD-NOMA shown in Fig. \ref{Fig.sub.2}, the BS maps signals of multiple users into a single RB by utilizing one row of the sparse matrix with dimension of $1 \times \frac{N}{2}$ and then transmits superposed signals to two users. It is observed that one row including multiple ``1"s denotes that the data streams from one user can be  spread into multiple layers, which can provide more flexibility for resource mapping. While for CD-NOMA, a user's data streams are directly mapped into multi-RB by sharing multiple spread sequences. As a further advance, the superposed signals of two users are formulated at the BS, which will be transmitted to the destination. Similar to uplink design, optimal design of the sparse matrix can further enhance the detection performance. At the receiver, MPA  is implemented to recover the desirable user's signals. From a practical perspective, computational  complexity at the reciter will grow exponentially with the number of users increasing. Hence, how to reduce the detection complexity for CD-NOMA should be taken into account in the 5G standardization process.

\subsection{User Association in NOMA Enabled HUDNs}
The distinct characteristics of the HUDNs with unified NOMA inevitably necessitate the redesign of user association algorithms. In contrast to the conventional user association approaches, on the one hand, the dense deployment of small cells introduce severe inter-interference as the neighbouring cells share the same RB. On the other hand, NOMA brings extra intra-interference from the same BS, hence making the user association design more challengeable.

\begin{figure}[!t]
    \begin{center}
        \includegraphics[width=4.6 in,  height= 1.4 in]{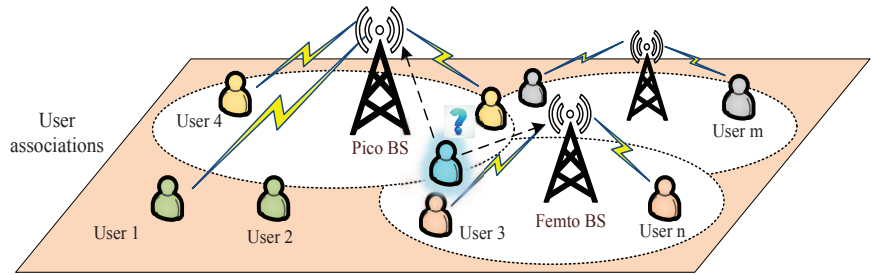}
        \caption{User association for NOMA enabled HUDNs.}
        \label{User association}
    \end{center}
\end{figure}

In order to address these two issues, we propose a flexible user association design for the unified NOMA enabled HUDNs, in which a NOMA user is allowed to access the BS of any tier in order to achieve the best coverage. As shown in Fig.~\ref{User association}, we take the PD-NOMA as a specific example. For simplicity, we consider that all BSs of HUDNs operate over the same orthogonal RB. Assuming that each user connects with one BS at most, while one BS can serve two users by adopting NOMA techniques. Particularly, we propose to associate users to BSs based on the maximum average power received at each NOMA user. In other words, each user not always access the nearest BS. It is allowed to access any tier BS. Such a user association scheme is fundamentally different from the conventional  approach, which associates users with the nearest BS and  may lead to the association of most users with small cells as their BSs are much closer to end users. As illustrated in Fig.~\ref{User association}, each BS has been associated with some users. When a new user joints the network, its association should be determined by considering the effects of both transmit power disparity of HUDNs and power sharing coefficients of NOMA users associated to the same BS.  Based on this flexible user association approach, network performance of NOMA enabled HUDNs has been investigated in~\cite{Liu2017HetNets}, which has analytically demonstrated that NOMA enabled HUNDs outperform the conventional OMA enabled one.

\subsection{Resource Allocation in NOMA Enabled HUDNs}
\begin{figure}[!t]
    \begin{center}
        \includegraphics[width=4.5 in,  height= 1.7 in]{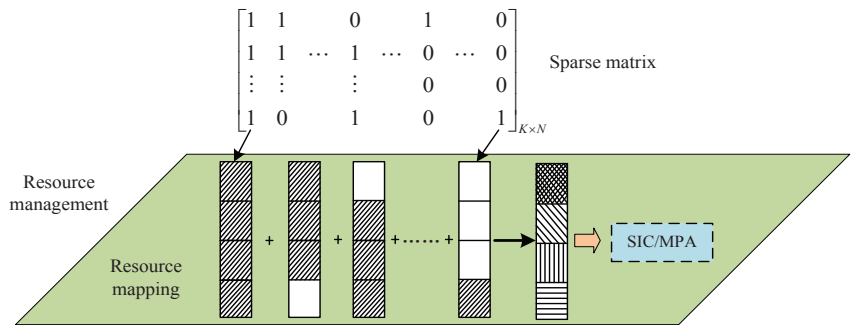}
        \caption{Resource allocation for NOMA enabled HUDNs.}
        \label{Resource_allocation}
    \end{center}
\end{figure}

Resource allocation is another significant aspect for designing NOMA enabled HUDNs. Note that the implementation of NOMA brings more sophisticated co-channel interference to  existing HUDNs, such distinct characteristics lead the resource allocation problems more challengeable.  Fig.~\ref{Resource_allocation} provides an illustration of resource allocation for our proposed unified NOMA enabled HUDNs framework. More particularly, date streams of different users can be spread over multiple RBs, where ``1" and ``0" denote whether there exists a resource mapping between the corresponding user and RB. More specifically, the shaded blocks refer to the RB occupied by users' data, which indicates a mapping. Technically, each user can select one  column from the sparse matrix randomly. However, to improve the detection performance, distant user prefers to select one column with larger column weight for resource allocation. While nearby user tends to select one column with smaller column weight. Furthermore, by multiplying a   power sharing coefficient with each column,  network performance  can be further enhanced. Finally, we employ MUD-based SIC/MPA to detect and output the information for the  desired user.

For resource allocation in NOMA enabled HUDNs, several problems should be jointly considered  for intelligently tackling the intra-BS and inter-BS interferences: i) the number of users to be allocated in the same RB; ii) which users should be allocated into which RB; and iii) the power sharing coefficient for each RB as well as for users sharing the same RB. For example, for single carrier system, the superposed signal of multiple users can be mapped into a single carrier over different power levels. The power allocation between NOMA users should be considered carefully. For multi-carrier system, the superposed signal can be mapped into multiple sub-carriers, where the NOMA users can select which sub-carrier to employ based on their requirements and then consider power allocation. All these problems can be addressed by properly designing the sparse matrix and its corresponding power sharing coefficients. Actually, due to the unique character of intra-BS interference brought by NOMA, resource allocation in NOMA enabled HUNDs becomes mix integer non-convex optimization problems, which usually tend to be NP-hard. Hence, efficient resource allocation algorithms are more than desired. Matching theory can be invoked as an effective approach for achieving good  tradeoff between system performance and computational complexity. With invoking matching theory, the authors in \cite{jingjing2016Hetnets} have proposed an effective resource allocation approach to show that NOMA enabled  HUNDs scheme is capable of achieving a higher sum rate compared to the OMA-enabled one.

\section{Case Study for  NOMA Enabled HUDNs}
In this section, we evaluate the unified NOMA enabled HUDNs  by simulations. For simplicity, we consider the uplink transmission of PD-NOMA. In the following, we provide two case studies to demonstrate user association and resource allocation in the unified NOMA enabled HUDNs, respectively.

\subsection{User Association in NOMA Enabled HUDNs}
In this study, we  illustrate how the density of small cells influences the user association in NOMA enabled HUDNs. Here, we consider  user association in the case where the proposed unified NOMA framework is applied in HUDNs based on a stochastic geometry model. More particularly, the locations of BSs and users follow homogeneous Poisson point processes. In the consider network,  macro cells employ massive MIMO and small cells adopt NOMA to support massive connectivity, and the maximum average received power approach is adopted to determine user association as aforementioned in section III. More details of the considered network configurations can be found in~\cite{Liu2017HetNets}.

\begin{figure}[!t]
    \begin{center}
        \includegraphics[width=3.7in, height=2.4in]{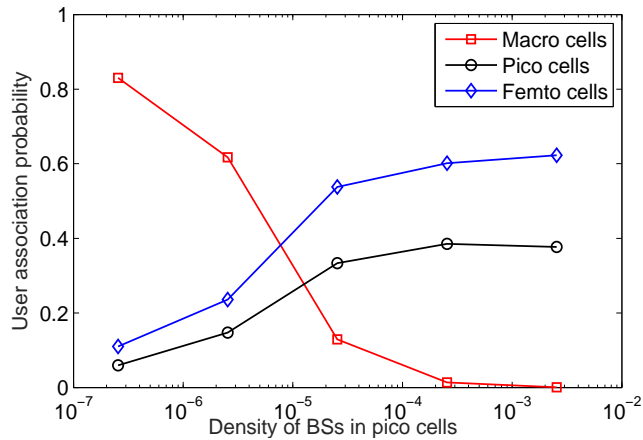}
        \caption{User association probability versus density of BSs in pico cells, $K=3$ tiers, number of antennas equipped in  macro cell BS $M=200$, data streams $N=15$, macro BS transmit power is ${P_1} = 40$ dBm, pico BS transmit power is ${P_2} = 30$ dBm, and femto BS transmit power is ${P_3} = 20$ dBm,  power sharing coefficients for NOMA users are $a_m=0.6$ and $a_n=0.4$, respectively, density of macro BS $\frac{1}{{2 \times {{500}^2}\pi }}$, density of small cell BS ${\lambda _{femto}}= 5\times {\lambda _{pico}}$.}
        \label{user asso}
    \end{center}
\end{figure}

Fig.~\ref{user asso} plots the user association probability of the unified NOMA enabled HUDNs with three layers, i.e., $K=3$. In this case, the density of macro BSs is fixed, while the densities of pico BSs and femto BSs vary correspondingly. It can be observed that NOMA users prefer macro cells when the density of small cells is relatively low. With the densification of small cells,  which is the case of HUDNs, NOMA users show a higher intention to be associated with BSs in pico and femto cells. It is also worth noting that NOMA users have a higher probability to be associated with BSs in femto cells even though BSs in pico cells transmit at a higher power level. This is caused by the dense deployment of femto cells, which also revels the effectiveness and  benefits of the NOMA enabled HUDNs.

\subsection{Resource Allocation in NOMA Enabled HUDNs}
In this study, we compare the performance of our unified NOMA enabled HUDNs with the conventional OMA enabled scheme in terms of both fairness and sum rates. We consider a HetNet with two tiers, in which the macro cell and small cells reuse the same set of RBs. In other words, we can refer to the small cells as the underlay tier. We allow each small cell BS to serve two users via NOMA by using the same RB. Our goal is to  maximize the sum rate of NOMA enabled small cell BSs via proper RB and power sharing coefficients schemes. More particularly, we adopt the matching theory for user allocation and sequential convex programming for power control \cite{jingjing2016Hetnets}.

\begin{figure} [!t]
\centering
\subfigure[Fairness comparison, transmit power at macro BSs is $43$ dBm, the transmit power at small cell BSs is $23$ dBm.]{
\label{fairness}
\includegraphics[width=3.15in,height=2.1in]{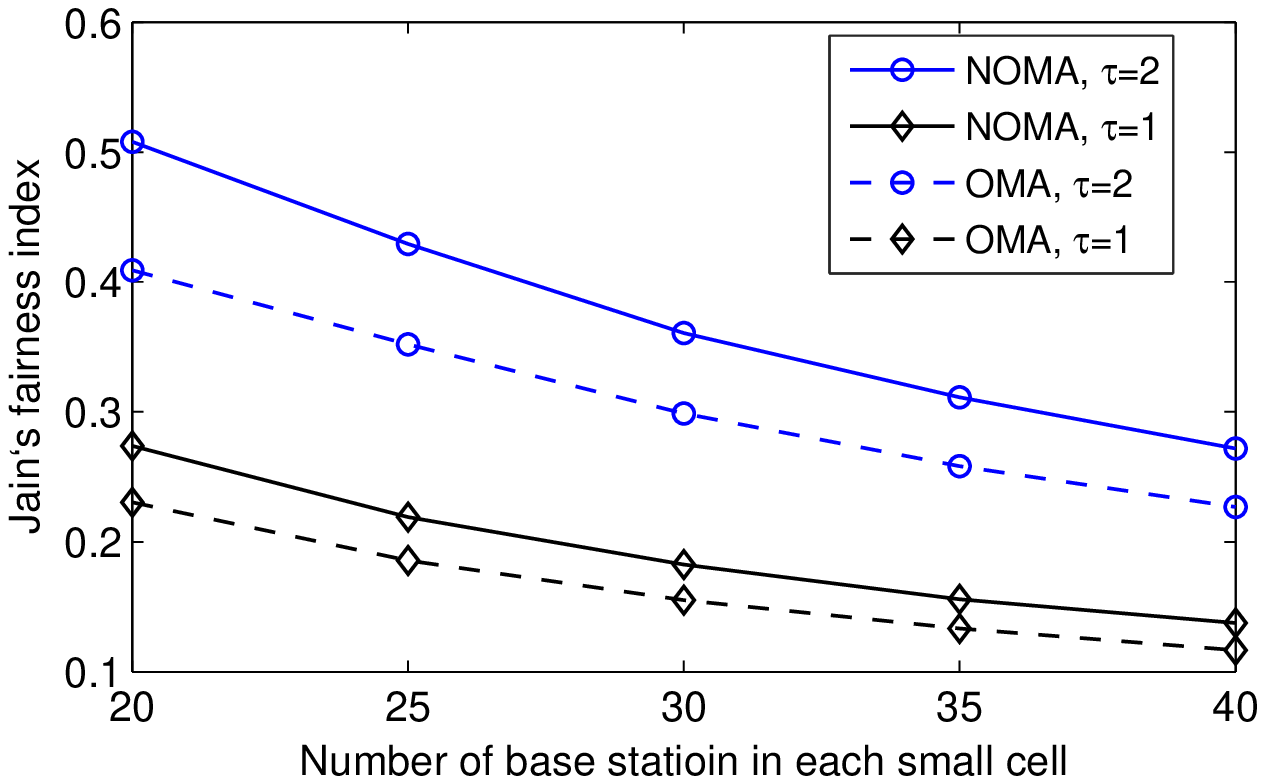}}
\subfigure[Sum rate comparison, maximum number of small cells occupying the same RB $\tau=2$.]{
\label{sum_rate}
\includegraphics[width=3.15in,height=2.1in]{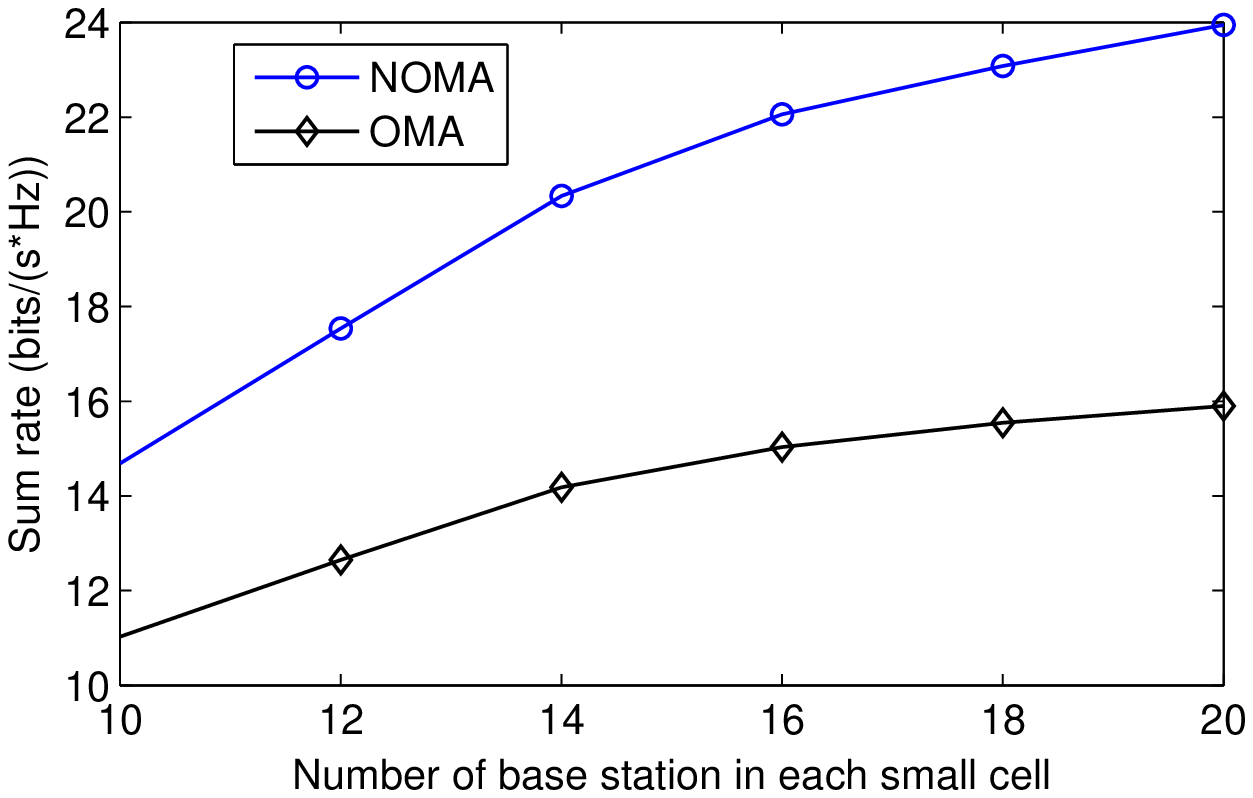}}
\caption{Resource allocation comparison between the unified NOMA enabled HUDNs and the OMA enabled scheme.}
\label{OMA_NOMA}
\end{figure}

Fig.~\ref{OMA_NOMA} plots the fairness and sum rate of resource allocation versus the total number of small cell BSs, respectively. Here, $\tau$ is the maximum number of small cell BSs occupying the same RB for restricting the co-channel interference.  Jain’s fairness \cite{fairness} index is adopted to evaluate the performance of the considered networks. We can observe that the fairness performance decreases with the number of small cell BSs in Fig.~\ref{fairness}. This is  because that large number of small cell BSs leads to more severe competition for limited spectrum resources. Consequently, more small cell BSs with poor channel conditions can not be accessed. We can also note that as $\tau$ increases, a higher fairness rate can be achieved. This is attributed to the fact that more small cell BSs can be multiplexed on each RB, which increases the multi-user diversity gain. It is also worth noting that the unfied NOMA enabled HUDNs have superior performance than the conventional OMA scheme both in terms of fairness and sum rate, which demonstrates the effectiveness the proposed structure.

\section{Research Challenges in NOMA Enabled HUDNs}
To support massive connectivity and enhance spectrum efficiency in 5G systems, the following  research problems should be addressed in the NOMA enabled HUNDs:

\begin{itemize}
\item \textbf{Energy efficiency in NOMA enabled HUNDs}: one of the potential applications of NOMA enabled HUNDs is IoT for smart cities, in which massive number of devices need to be connected. The unified NOMA enabled HUNDs provide a practical infrastructure to offer massive access opportunities for such  large number of devices, especially for the cases that each device only needs to send a small amount of data periodically. However, these devices  are normally restricted to power consumption as they are powered by battery. In order to extend the battery lifetime of these IoT devices, i.e., devices are expected to keep live for ten years,  the energy efficiency is under investigated. Particularly, higher data transmission rate  results in shorter airtime, however, the power consumption for data transmission becomes higher. With lower power consumption, the achieved transmission rate becomes lower, which extends the airtime. Therefore, the tradeoff between data rate and airtime should be considered to maximize battery lifetime of devices.
 \item \textbf{Big Data aided Adaptive NOMA in HUNDs}: IoT devices normally have limited processing capability, while some devices, i.e., mobile phones, are capable of performing more complex tasks. Meanwhile, it is noted that different NOMA schemes requires different complexity levels at the user side. For instance, SIC receiver is relatively simple, which makes PD-NOMA more suitable for IoT devices.  Therefore, a software-defined NOMA network architecture is desired to achieve adaptive NOMA with awareness on complexity to support different user scenarios.  Machine learning can be invoked to predict the data traffic for different user scenarios. Adaptive MA technique can be categorized into two types, including  pre-settings and real-time settings. Pre-settings refer to assign the MA technique according to historical social media information, such as the number of users within a given area in several months or one year. The real-time settings adjust the MA  technique based  on the real-time feedback from social media.

  \item \textbf{Testbed for NOMA enabled HUNDs}: even though extensive research has been carried out on the performance analyses and algorithm designs for NOMA enabled HUNDs, there is still a large gap to the real implementation of NOMA in HUNDs. For the proof of concept, a testbed is more than desired to demonstrate the effectiveness of the proposed unified NOMA enabled HUNDs. More particularly, the idea of software defined radio  (SDR) can be adopted to enable BSs to select the proper NOMA technique smartly according to different application scenarios.

\end{itemize}

\section{Conclusions}
This article has envisioned NOMA enabled HUDNs as a promising solution to support massive connectivity in 5G systems. Instead of focusing on specific NOMA techniques individually, we have proposed a unified NOMA framework. Moreover, we have investigated the application of the proposed unified NOMA framework in HUDNs. We have further explored the critical challenges on user association and resource allocation in NOMA enabled HUDNs,  as  both the dense deployment of small cells and the non-orthogonality in resource sharing bring severe interference. Additionally, we have carried out related case studies, which have provided important insights for the future design of NOMA enabled HUDNs to support massive connectivity in 5G systems.

\bibliographystyle{IEEEtran}
%\bibliography{mybib}

 \end{document}